\documentclass[preprint,12pt]{elsarticle}

\usepackage{amssymb}
\usepackage{amsmath}
\usepackage{url}

\journal{Physica A}

\begin{document}

\begin{frontmatter}



\title{Separating causal texts from non-causal texts is not evidence of
entropy production in language models} 


\author[icr]{Y-h. Taguchi\corref{cor1}\fnref{fn1}}
\ead{tag@granular.com}
\cortext[cor1]{Corresponding author}
\fntext[fn1]{Permanent address: Department of Physics, Chuo University,
1-13-27 Kasuga, Bunkyo-ku, Tokyo 112-8551, Japan}

\affiliation[icr]{organization={Bioinformatics Center, Institute for Chemical Research, Kyoto University},
            addressline={Gokasho, Uji},
            city={Kyoto},
            postcode={611-0011},
            country={Japan}}

\begin{abstract}A recent study defines an entropy production for autoregressive language
models.
The block-level entropy production compares a written text with the same
sentences in the reversed order.
The study reports larger values of the block-level entropy production for
causal texts than for non-causal texts.
This paper evaluates the same quantity, divided by the number of tokens, over
all 24 arrangements of the four sentences of each written text.
The tokenization of the study adds a leading space and reproduces the
published values.
With the tokenization of the study, the causal order accounts for 71\% to 93\%
of the rank of the written arrangement for five autoregressive models.
The separation holds also on two new corpora in which causal graphs, fixed
before generation and checked by judge models, define the causal texts.
The leading space is a choice outside the definition of the quantity.
Without the leading space, the share of the causal order drops to 0\% to 21\%,
and an earlier version of this paper reached the opposite conclusion for the
same reason.
Two masked diffusion language models give a second quantity, which compares
left-to-right unmasking and right-to-left unmasking of the same token sequence
and has a zero on each token sequence.
The second quantity separates causal texts from non-causal texts only weakly,
whereas the block-level quantity of the same models separates causal texts
from non-causal texts clearly.
The ability of a quantity to separate causal texts from non-causal texts is
therefore not evidence that the quantity is an entropy production.
\end{abstract}


\end{frontmatter}



\section{Introduction}
\label{sec:introduction}


Ref.~\cite{sagawa2026} develops stochastic thermodynamics for autoregressive
generative models and defines an entropy production of a token sequence.
This paper calls the autoregressive generative models of
Ref.~\cite{sagawa2026} autoregressive models.
Ref.~\cite{sagawa2026} evaluates the block-level entropy production, which
reverses the order of the sentences, on externally prepared
texts~\cite[Sec.~V~B]{sagawa2026}.
Ref.~\cite{sagawa2026} reports larger values of the block-level entropy
production for causal texts than for non-causal texts.
In causal texts, each sentence states a consequence of the preceding sentence.
In non-causal texts, the order of the sentences carries no causal relation.
Ref.~\cite{sagawa2026} regards the evaluation as a first step toward probing
the temporal order or the causal order of the events that a text
describes~\cite[Sec.~I]{sagawa2026}.

A written text consists of four sentences, named sentences A, B, C and D.
The symbol $x$ denotes the token sequence for the written arrangement, and
Section~\ref{sec:tokenization} defines how $x$ is constructed.
The token sequence $\tilde{x}$ is obtained from $x$ by reversing the order of
the four sentence token sequences, so that $\tilde{x}$ corresponds to the
reversed arrangement D C B A.
An autoregressive model assigns a log-probability to the token sequence $x$
and a log-probability to the token sequence $\tilde{x}$.
The quantity
\begin{equation}
\Delta_{\mathrm{block}} \;=\; \frac{\log P(x) - \log P(\tilde{x})}{L}
\label{eq:dblock}
\end{equation}
is the difference of the two log-probabilities, divided by the number $L$ of
tokens in $x$.
For the written arrangement, $L\,\Delta_{\mathrm{block}}$ is the block-level
entropy production of Ref.~\cite{sagawa2026}.

Ref.~\cite{sagawa2026} states two open problems about the comparison.
The first open problem is that a generation prompt, and not a formal
criterion, defines which texts are causal texts~\cite[Sec.~V~B]{sagawa2026}.
The second open problem is that $\Delta_{\mathrm{block}}$ may mix causal order
with temporal order and with other properties of the order of the
sentences~\cite[Sec.~VIII]{sagawa2026}.

This paper asks two questions.
The first question is how well $\Delta_{\mathrm{block}}$ separates causal texts
from non-causal texts when the separation is measured within each written
text.
The second question is whether the ability of $\Delta_{\mathrm{block}}$ to
separate causal texts from non-causal texts supports the reading of
$\Delta_{\mathrm{block}}$ as an entropy production.

The answer to the first question uses the $4! = 24$ arrangements of the four
sentences of each written text.
For each written text, the midrank is the rank of the value of
$\Delta_{\mathrm{block}}$ for the written arrangement among the 24 values of
$\Delta_{\mathrm{block}}$ for the 24 arrangements.
The midrank is scaled to the range from 0 to 1.
A midrank of 0.5 is the middle of the 24 values.
The causal share is the part of the distance of the midrank from 0.5 that
separates causal texts from non-causal texts.

The first open problem of Ref.~\cite{sagawa2026} concerns how causal texts are
defined.
In the corpora of Ref.~\cite{sagawa2026}, a causal text is a written text that
a large language model generated when the generation prompt asked for causally ordered
sentences.
This paper constructs two new corpora, the Claude--Gemini corpus and the
ChatGPT corpus (Section~\ref{sec:corpus}).
Claude and Gemini generated the written texts of the Claude--Gemini corpus,
and ChatGPT generated the written texts of the ChatGPT corpus.
For each written text of the Claude--Gemini corpus and the ChatGPT corpus, a
causal graph over sentences A, B, C and D is fixed before generation.
A causal text has the chain graph A$\to$B$\to$C$\to$D, and a non-causal text
has the empty graph.
After generation, two judge models checked that the causal graph of every
written text agrees with the causal graph fixed before generation.
The two judge models come from companies whose large language models did not
generate the written text.

For the 24 arrangements, the value of $\Delta_{\mathrm{block}}$ depends on how
the token sequences $x$ and $\tilde{x}$ are constructed.
Procedure 2-T tokenizes the written text and arranges the sentence token
sequences.
Procedure 2-S arranges the sentence character strings and tokenizes the
arranged sentence character string.
This paper compares three settings.
Setting 1 uses procedure 2-T with a leading space, as Ref.~\cite{sagawa2026}
does.
Setting 2A uses procedure 2-T without a leading space.
Setting 2B uses procedure 2-S without a leading space.
With a leading space, procedure 2-T and procedure 2-S produce the same token
sequence for every arrangement.
In setting 2A, procedure 2-T produces a mismatch with the tokenizer output in
every arrangement that does not start with sentence A.
In setting 2B, procedure 2-S produces no mismatch with the tokenizer output.
The first version of this paper~\cite{taguchi2026v1} used setting 2A, and the
mismatch with the tokenizer output reversed the conclusion of the first
version of this paper.

The answer to the second question uses masked language models.
A masked language model (MLM) computes the log-probability of a token sequence
by unmasking the tokens one at a time.
The computation starts from the token sequence in which all tokens are masked.
Under left-to-right unmasking, the MLM unmasks the tokens from the first token
to the last token.
Under right-to-left unmasking, the MLM unmasks the tokens from the last token
to the first token.
The two unmasking orders give two log-probabilities of the same token
sequence.
The quantity $\sigma$ compares the two log-probabilities of one token sequence.
The value of $\sigma$ for the written arrangement needs no arrangement of
sentences, and therefore does not depend on procedure 2-T or procedure 2-S.
For an MLM whose conditional probabilities come from one joint distribution,
$\sigma$ is zero for every token sequence.

The results are the following.
\begin{enumerate}
\item
With setting 1, $\Delta_{\mathrm{block}}$ of the written arrangement reproduces
the values of Ref.~\cite{sagawa2026} for all 1000 written texts of the corpora
of Ref.~\cite{sagawa2026} (Section~\ref{sec:method}).
\item
Five autoregressive models computed $\Delta_{\mathrm{block}}$ with setting 1 for
the written texts of the corpora of Ref.~\cite{sagawa2026}.
The causal share is 71\% to 93\%.
The causal share is 64\% to 100\% on the Claude--Gemini corpus and 62\% to
127\% on the ChatGPT corpus (Section~\ref{sec:delta}).
\item
With setting 2A, the causal share drops to 0\% to 21\%.
With setting 2B, the causal share stays close to the causal share with
setting 1
(Section~\ref{sec:settings}).
\item
The value of $\sigma$ separates causal texts from non-causal texts only weakly
for three MLMs (Section~\ref{sec:sigma}).
\item
For two MLMs, LLaDA and iLLaDA, $\Delta_{\mathrm{block}}$ computed with
left-to-right unmasking separates causal texts from non-causal texts as
clearly as $\Delta_{\mathrm{block}}$ for the autoregressive models
(Section~\ref{sec:mlmdelta}).
\end{enumerate}

The fifth result answers the second question.
For an MLM, left-to-right unmasking and right-to-left unmasking are two
processes that the MLM actually runs on the same token sequence, in opposite
unmasking orders.
For an MLM whose conditional probabilities come from one joint distribution,
$\sigma$ is zero for every token sequence.
The quantity $\sigma$ therefore has a zero on each token sequence, as an
entropy production has a zero for a reversible process.
The quantity $\Delta_{\mathrm{block}}$ compares the token sequences of two
arrangements, and $\Delta_{\mathrm{block}}$ has no zero that holds on each
written text.
With respect to the zero, $\sigma$ is closer to an entropy production than
$\Delta_{\mathrm{block}}$.
However, $\sigma$ separates causal texts from non-causal texts only weakly,
whereas $\Delta_{\mathrm{block}}$ separates causal texts from non-causal texts
clearly.
The quantity closer to an entropy production loses the separation, and the
quantity that keeps the separation is a comparison between two token
sequences.
The ability of a quantity to separate causal texts from non-causal texts is
therefore not evidence that the quantity is an entropy production.

\section{Methods}
\label{sec:method}

\subsection{Quantities}
\label{sec:quantities}


\paragraph{Arrangements}
A written text consists of four sentences, named sentences A, B, C and D in
the order in which the four sentences appear in the written text.
An arrangement is an order of sentences A, B, C and D.
The symbol $\pi$ denotes an arrangement, and the symbol $\bar{\pi}$ denotes the
reversed arrangement of $\pi$.
For example, the reversed arrangement of the arrangement B A D C is the
arrangement C D A B.
The written arrangement is A B C D.
Each written text has $4! = 24$ arrangements.

\paragraph{Token sequences}
For an arrangement $\pi$, the token sequence $x_\pi$ is a token sequence of the
four sentences in the order $\pi$.
Section~\ref{sec:tokenization} defines how the token sequence $x_\pi$ is
constructed.
The number $L$ is the number of tokens in $x_\pi$ when $\pi$ is the written
arrangement.
The number of tokens in $x_\pi$ when $\pi$ is another arrangement can differ
from $L$ (Section~\ref{sec:tokenization}), and $L$ divides the quantities of
every arrangement of one written text.

\paragraph{Log-probabilities}
An autoregressive model assigns to a token sequence
$x = (x_1, \ldots, x_n)$ the log-probability
\begin{equation}
\log P(x) \;=\; \sum_{k=1}^{n} \log P(x_k \mid x_1, \ldots, x_{k-1}).
\label{eq:arlogp}
\end{equation}
A masked language model (MLM) predicts masked tokens of a token sequence from
the tokens that are not masked.
An MLM computes a log-probability of a token sequence $x = (x_1, \ldots, x_n)$
by unmasking the $n$ tokens one at a time.
The computation follows the generation process of an MLM, for example LLaDA.
In the generation process, the MLM starts from a token sequence in which all
tokens are masked, and the MLM unmasks the tokens step by step.
The unmasking order is the order in which the $n$ positions of $x$ are
unmasked.
To compute the log-probability of $x$, the MLM starts from the token sequence
in which all $n$ tokens are masked.
At step $k$, the MLM predicts the token at the $k$-th position of the unmasking
order from the tokens already unmasked.
The token at the $k$-th position is then unmasked and set to the actual token
of $x$.
The log-probability of $x$ under the unmasking order is the sum of the $n$
logarithms of the predicted probabilities of the actual tokens.
Left-to-right unmasking uses the unmasking order $1, 2, \ldots, n$, and gives
the log-probability $\log P_{\mathrm{l2r}}(x)$.
Right-to-left unmasking uses the unmasking order $n, n-1, \ldots, 1$, and
gives the log-probability $\log P_{\mathrm{r2l}}(x)$.
Two different unmasking orders can give two different log-probabilities of the
same token sequence $x$.

\paragraph{The quantity $\Delta_{\mathrm{block}}$}
For an arrangement $\pi$,
\begin{equation}
\Delta_{\mathrm{block}}(\pi) \;=\;
\frac{\log P(x_\pi) - \log P(x_{\bar{\pi}})}{L}.
\label{eq:dblock-pi}
\end{equation}
For the written arrangement, $L\,\Delta_{\mathrm{block}}$ is the block-level
entropy production of Ref.~\cite{sagawa2026} (Section~\ref{sec:reproduction}).
The definition~\eqref{eq:dblock-pi} gives
$\Delta_{\mathrm{block}}(\bar{\pi}) = -\Delta_{\mathrm{block}}(\pi)$
for every arrangement $\pi$.
The 24 values of $\Delta_{\mathrm{block}}$ for one written text are therefore
symmetric about zero.
For an autoregressive model, $\log P$ in Eq.~\eqref{eq:dblock-pi} is the
log-probability~\eqref{eq:arlogp}.
For an MLM, $\log P$ in Eq.~\eqref{eq:dblock-pi} is $\log P_{\mathrm{l2r}}$:
\begin{equation}
\Delta_{\mathrm{block}}(\pi) \;=\;
\frac{\log P_{\mathrm{l2r}}(x_\pi) - \log P_{\mathrm{l2r}}(x_{\bar{\pi}})}{L}.
\label{eq:dblock-l2r}
\end{equation}
Left-to-right unmasking corresponds to the left-to-right generation of an
autoregressive model.

\paragraph{The quantity $\sigma$}
For an MLM and a token sequence $x$,
\begin{equation}
\sigma(x) \;=\; \frac{\log P_{\mathrm{l2r}}(x) - \log P_{\mathrm{r2l}}(x)}{L}.
\label{eq:sigma}
\end{equation}
The quantity $\sigma$ compares two log-probabilities of the same token sequence
$x$.
For an MLM whose conditional probabilities come from one joint distribution
$P$, both $\log P_{\mathrm{l2r}}(x)$ and $\log P_{\mathrm{r2l}}(x)$ equal
$\log P(x)$ by the chain rule of probability.
The quantity $\sigma$ therefore has a zero on each token sequence.
A nonzero value of $\sigma$ shows that the conditional probabilities of the MLM
do not come from one joint distribution.
For each written text, the 24 values $\sigma(x_\pi)$ for the 24 arrangements
are computed.

\paragraph{Midrank}
For each written text, the midrank of the written arrangement is
\begin{equation}
m \;=\; \frac{n_{<} + \tfrac{1}{2}\,n_{=}}{24}.
\label{eq:midrank}
\end{equation}
The number $n_{<}$ counts the arrangements whose value of
$\Delta_{\mathrm{block}}$ is smaller than the value of
$\Delta_{\mathrm{block}}$ for the written arrangement.
The number $n_{=}$ counts the arrangements whose value of
$\Delta_{\mathrm{block}}$ equals the value of $\Delta_{\mathrm{block}}$ for
the written arrangement.
The written arrangement itself is one of the arrangements counted by $n_{=}$,
so that $n_{=} \geq 1$.
A midrank of 0.5 is the middle of the 24 values.
If the 24 values of $\Delta_{\mathrm{block}}$ are all equal, the midrank is
0.5.

\paragraph{Causal share}
For each corpus and each model, the median midrank of causal texts and the
median midrank of non-causal texts are computed.
In non-causal texts, the order of the sentences carries no causal relation.
The median midrank of non-causal texts minus 0.5 therefore measures how far
the written arrangement is ranked higher than the middle.
The reason for the higher rank is only that the large language model that
generated the written text produced the sentences in the written arrangement.
This paper calls the median midrank of non-causal texts minus 0.5 the authored
component.
The median midrank of causal texts minus the median midrank of non-causal
texts measures the part that the causal order adds.
This paper calls the median midrank of causal texts minus the median midrank
of non-causal texts the causal component.
The causal share is the causal component divided by the sum of the authored
component and the causal component.
For the five corpora of Ref.~\cite{sagawa2026}, the authored component and the
causal component are computed for each corpus, and the medians over the five
corpora enter the causal share.
The causal share exceeds 100\% when the authored component is negative.

\paragraph{Causal share in $z$}
For each written text, $z$ is the value of $\Delta_{\mathrm{block}}$ for the
written arrangement divided by the standard deviation of the 24 values of
$\Delta_{\mathrm{block}}$.
The causal share in $z$ is the causal share computed with $z$ in place of the
midrank and with 0 in place of 0.5.
The value $z = 0$ corresponds to a midrank of 0.5, because the 24 values of
$\Delta_{\mathrm{block}}$ are symmetric about zero.

\paragraph{Comparison between causal texts and non-causal texts}
The rank-biserial correlation $r$ compares a quantity between causal texts and
non-causal texts, as in Ref.~\cite{sagawa2026}.
A positive value of $r$ means that the quantity tends to be larger for causal
texts.
The $p$-value comes from the two-sided Mann--Whitney $U$ test.

\subsection{Construction of token sequences}
\label{sec:tokenization}


\paragraph{Sentences}
Each written text of the corpora of Ref.~\cite{sagawa2026}, the Claude--Gemini
corpus and the ChatGPT corpus consists of four sentences.
In every written text, exactly one space follows each sentence-final
``.'', ``!'' or ``?'', except the last one, and no space starts or ends the
written text.
The token sequence of the written text is the output of the tokenizer for the
written text.
Each token is converted to a character string, and the spaces at the end of
the character string are removed.
A sentence-final token is a token whose resulting character string ends with
``.'', ``!'' or ``?''.
A sentence boundary in a token sequence lies directly after a sentence-final
token.
The token sequence of the written text, or the token sequence of the written
text with a leading space, splits at the sentence boundaries into four parts.
The four parts are the four sentence token sequences.
A sentence character string is obtained from a sentence token sequence in two
steps.
First, the sentence token sequence is converted to a character string.
Second, the spaces at the start and at the end of the character string are
removed.
The written text equals the four sentence character strings in the written
arrangement, joined with one space.

\paragraph{Leading space}
The tokenizers of gpt2, Qwen3, LLaDA and iLLaDA include the space before a word
in the token of the word.
A word after a space and the same word at the start of a character string are
therefore different tokens.
In the written text, no space precedes sentence A, because sentence A is at
the start of the written text.
In the written text, one space precedes each of sentences B, C and D.
Ref.~\cite{sagawa2026} adds one space at the start of each written text
before tokenization~\cite[App.~B~4 and App.~C]{sagawa2026}.
Ref.~\cite{sagawa2026} states that the leading space preserves the space
between sentences when the sentence token sequences are
rearranged~\cite[App.~B~4 and App.~C]{sagawa2026}.
A computation of this paper with a leading space adds one space at the start
of every character string before tokenization.
A computation of this paper without a leading space adds no space.

\paragraph{Two procedures}
Procedure 2-T tokenizes the written text, splits the token sequence into the
four sentence token sequences, and joins the four sentence token sequences in
the arrangement $\pi$.
Procedure 2-S joins the four sentence character strings in the arrangement
$\pi$ with one space, and tokenizes the arranged sentence character string.
With a leading space, procedure 2-T adds the space to the written text, and
procedure 2-S adds the space to the arranged sentence character string.
For the written arrangement, procedure 2-T and procedure 2-S give the same token
sequence.

\paragraph{Mismatch with the tokenizer output}
For an arrangement $\pi$, a mismatch with the tokenizer output is a position at
which the token sequence of procedure 2-T differs from the token sequence of
procedure 2-S.
Without a leading space, every arrangement that does not start with sentence A
has two mismatches with the tokenizer output.
The first mismatch is the first token of the token sequence.
Procedure 2-T keeps the token that begins with a space character, because the
first sentence of the arrangement followed a space in the written text.
Procedure 2-S gives the token that does not begin with a space character,
because the sentence starts the arranged sentence character string.
The second mismatch is the first token of sentence A.
Procedure 2-T keeps the token that does not begin with a space character,
because sentence A started the written text.
Procedure 2-S gives the token that begins with a space character, because
sentence A follows a space in the arranged sentence character string.
With a leading space, every sentence follows a space in both procedures, and
the two procedures give the same token sequence for every arrangement.
Table~\ref{tab:mismatch} counts the arrangements in which a mismatch with the
tokenizer output occurs.

\begin{table}[t]
\centering
\caption{Number of arrangements in which a mismatch with the tokenizer output
occurs, over all written texts of each corpus.
Each written text has 6 arrangements that start with sentence A and 18
arrangements that do not start with sentence A, 24 arrangements in all.
The column ``Number'' gives the number of written texts times 6, 18 or 24.
The tokenizers of gpt2, Qwen3, LLaDA and iLLaDA give the same counts.
The tokenizer of bert-base gives no mismatch with or without a leading space.}
\label{tab:mismatch}
\small
\begin{tabular}{lrrr}
\hline
 & & \multicolumn{2}{c}{Arrangements with a mismatch} \\
Arrangements & Number & without & with \\
 & & a leading space & a leading space \\
\hline
\multicolumn{4}{l}{Ref.~\cite{sagawa2026}, each corpus (200 written texts)} \\
\quad starting with A & $200 \times 6 = 1200$ & 0 & 0 \\
\quad not starting with A & $200 \times 18 = 3600$ & 3600 & 0 \\
\quad all & $200 \times 24 = 4800$ & 3600 & 0 \\
\hline
\multicolumn{4}{l}{Claude--Gemini corpus (688 written texts)} \\
\quad starting with A & $688 \times 6 = 4128$ & 0 & 0 \\
\quad not starting with A & $688 \times 18 = 12384$ & 12384 & 0 \\
\quad all & $688 \times 24 = 16512$ & 12384 & 0 \\
\hline
\multicolumn{4}{l}{ChatGPT corpus (159 written texts)} \\
\quad starting with A & $159 \times 6 = 954$ & 0 & 0 \\
\quad not starting with A & $159 \times 18 = 2862$ & 2862 & 0 \\
\quad all & $159 \times 24 = 3816$ & 2862 & 0 \\
\hline
\end{tabular}
\end{table}

\paragraph{Three settings}
The definition of the block-level entropy production in
Ref.~\cite{sagawa2026} requires that $x_{\bar{\pi}}$ consist of the sentence
token sequences of $x_\pi$ in the reversed order~\cite[Secs.~IV~D and V]{sagawa2026}.
Procedure 2-T satisfies the requirement for every arrangement.
Without a leading space, procedure 2-S does not satisfy the requirement,
because the first token of sentence A depends on the position of sentence A.
This paper compares three settings for $\Delta_{\mathrm{block}}$.
Setting 1 uses procedure 2-T with a leading space, as
Ref.~\cite{sagawa2026} does.
Setting 2A uses procedure 2-T without a leading space.
Setting 2B uses procedure 2-S without a leading space.
Setting 1 satisfies the requirement and has no mismatch with the tokenizer
output.
Setting 2A satisfies the requirement and has mismatches with the tokenizer
output.
Setting 2B has no mismatch with the tokenizer output and does not satisfy the
requirement.
The values of $\sigma$ and the values of $\Delta_{\mathrm{block}}$ for MLMs use
setting 1.

\paragraph{Implementation of setting 1}
The computations of this paper for setting 1 run procedure 2-S with a leading
space.
With a leading space, procedure 2-S gives the same token sequence as procedure
2-T for every arrangement of every written text (Table~\ref{tab:mismatch}).
The computations for setting 1 therefore use the token sequences of setting 1.

\paragraph{Normalization}
In setting 1 and setting 2A, every token sequence $x_\pi$ of one written text
has $L$ tokens.
In setting 2B, the number of tokens in $x_\pi$ can differ from $L$.
In every setting, $L$ divides $\Delta_{\mathrm{block}}(\pi)$ and
$\sigma(x_\pi)$ for every arrangement $\pi$ of one written text.
The common divisor keeps $\Delta_{\mathrm{block}}(\bar{\pi}) =
-\Delta_{\mathrm{block}}(\pi)$.
The midrank and $z$ of one written text do not depend on the common divisor.

\paragraph{Initial context token}
An autoregressive model computes $\log P(x)$ with a fixed initial context token
before $x$, as in Ref.~\cite{sagawa2026}.
The fixed initial context token is the beginning-of-sequence token of the
tokenizer, or the end-of-sequence token when the tokenizer has no
beginning-of-sequence token.
For the four gpt2 models and Qwen3-4B, the fixed initial context token is
\texttt{<|endoftext|>}, as in Ref.~\cite[App.~B~1 and App.~C]{sagawa2026}.

\subsection{Corpora}
\label{sec:corpus}


\paragraph{Corpora of Ref.~\cite{sagawa2026}}
Five large language models generated the five corpora of
Ref.~\cite{sagawa2026}: Claude Opus 4.6, Claude Fable 5, Gemini 3.1 Pro,
GPT-5.4 Pro and GPT-5.6 Pro~\cite[Sec.~V~B]{sagawa2026}.
Each large language model received the same generation prompt and generated
100 causal texts and 100 non-causal texts, each with four sentences.
The generation prompt defines causal texts as short narratives in which the
sentences describe a temporally ordered sequence of events.
The generation prompt defines non-causal texts as collections of independent
statements whose order has no temporal or causal implication.
In every written text, the four sentences begin with the same noun phrase and
use the simple past tense, without pronouns, demonstratives or connectives
between sentences~\cite[Sec.~V~B]{sagawa2026}.

\paragraph{Causal graphs}
This paper constructs the Claude--Gemini corpus and the ChatGPT corpus from
causal graphs fixed before generation.
For a causal text, the causal graph is the chain graph
A$\to$B$\to$C$\to$D.
In the chain graph, the event of sentence A directly causes the event of
sentence B.
The event of sentence B directly causes the event of sentence C.
The event of sentence C directly causes the event of sentence D.
The chain graph has no other direct causal link.
For a non-causal text, the causal graph is the empty graph.
In the empty graph, no event causes, enables or prevents another event.

\paragraph{Generation}
A large language model received a generation prompt that states the causal
graph of each written text to generate.
The generation prompt requires the following properties of every written text.
\begin{enumerate}
\item The four sentences concern one entity, and a noun phrase of two to four
words for the entity appears unchanged in every sentence.
\item Each sentence can be understood without the other three sentences.
No sentence contains a pronoun, a connective, a time expression or a phrase
that refers to an event of another sentence.
\item No sentence states causation with a word from a list that includes
``cause'', ``lead'', ``result'', ``due'' and ``trigger''.
\item Each sentence is in the simple past tense, has 8 to 20 words and ends
with one period.
\end{enumerate}
For a non-causal text, the generation prompt also requires that four
different, unrelated agents act on or near the entity.
For a non-causal text, no event may change the condition of the entity.
The generated written texts were checked against the four properties before
judging.

\paragraph{Judging}
Two judge models judged every generated written text.
Each judge model received the four sentences in a random order.
Each judge model returned every pair of sentences in which the event of one
sentence directly causes the event of the other sentence.
A written text was accepted when, for both judge models, the transitive closure
of the returned causal links equals the transitive closure of the causal graph
fixed before generation.
The judge models came from companies whose large language models did not
generate the written text.
Table~\ref{tab:corpora} gives the large language models and the numbers of
written texts.

\begin{table}[t]
\centering
\caption{Construction of the Claude--Gemini corpus and the ChatGPT corpus
(November 2026).
The Claude--Gemini corpus consists of 18 batches.
Each batch has 20 written texts with the chain graph and 20 written texts with
the empty graph, and Claude Opus 5.5 and Gemini 4 Neon generated alternate
batches.
ChatGPT-6.2 Astra and Gemini 4 Neon judged the batches that Claude Opus 5.5
generated.
ChatGPT-6.2 Astra and Claude Opus 5.5 judged the batches that Gemini 4 Neon
generated.
The ChatGPT corpus consists of 4 batches of the same form, which ChatGPT-6.2
Astra generated and Claude Opus 5.5 and Gemini 4 Neon judged.}
\label{tab:corpora}
\begin{tabular}{lrr}
\hline
 & \multicolumn{2}{c}{Accepted / generated} \\
Generator & Chain graph & Empty graph \\
\hline
\multicolumn{3}{l}{Claude--Gemini corpus} \\
\quad Claude Opus 5.5 & 175 / 180 & 171 / 180 \\
\quad Gemini 4 Neon & 163 / 180 & 179 / 180 \\
\quad total & 338 / 360 & 350 / 360 \\
\hline
\multicolumn{3}{l}{ChatGPT corpus} \\
\quad ChatGPT-6.2 Astra & 79 / 80 & 80 / 80 \\
\hline
\end{tabular}
\end{table}

\paragraph{Pilot runs with open-weight models}
Before the construction of the Claude--Gemini corpus and the ChatGPT corpus,
five pilot runs used open-weight models on a CPU.
In the pilot runs, Qwen2.5-7B-Instruct generated the written texts.
Llama-3.1-8B-Instruct and Gemma-2-9B-it judged the written texts of the first
pilot run, and Gemma-2-9B-it and Phi-4 judged the written texts of the other
four pilot runs.
The generation prompt and the checks were revised between the pilot runs.
Of 38 generated written texts, 19 with the chain graph and 19 with the empty
graph, 27 passed the checks and 10 were accepted by both judge models.
The 10 accepted written texts were 4 of 19 with the chain graph and 6 of 19
with the empty graph.
The generated chains were often ambiguous or out of order, and the judge models
often returned causal links between independent events.
The construction of the Claude--Gemini corpus and the ChatGPT corpus therefore
used the large language models of Table~\ref{tab:corpora}.

\paragraph{Written texts}
The accepted written texts form the two corpora.
The Claude--Gemini corpus has 338 causal texts and 350 non-causal texts.
The ChatGPT corpus has 79 causal texts and 80 non-causal texts.
In the written arrangement A B C D, the events of a causal text appear in the
causal order of the chain graph.

\subsection{Models}
\label{sec:models}


\paragraph{Autoregressive models}
Five autoregressive models compute $\Delta_{\mathrm{block}}$: gpt2,
gpt2-medium, gpt2-large and gpt2-xl~\cite{gpt2}, and Qwen3-4B-Base~\cite{qwen3}
(Table~\ref{tab:models}).
Ref.~\cite{sagawa2026} uses gpt2 and Qwen3-4B-Base.
The four gpt2 models differ only in size, from 124 million to 1.5 billion
parameters.
All five autoregressive models compute log-probabilities in 32-bit floating
point.

\paragraph{Masked language models}
Three MLMs compute $\sigma$ and $\Delta_{\mathrm{block}}$: LLaDA-8B-Base~\cite{llada},
iLLaDA-8B-Base~\cite{illada} and bert-base-uncased~\cite{bert}.
The three MLMs compute $\Delta_{\mathrm{block}}$ with left-to-right unmasking
(Eq.~\eqref{eq:dblock-l2r}).
The three MLMs differ in the training mask ratio.

\paragraph{LLaDA-8B-Base and iLLaDA-8B-Base}
LLaDA-8B-Base and iLLaDA-8B-Base are masked diffusion language models with 8
billion parameters.
The training of LLaDA-8B-Base and iLLaDA-8B-Base masks a fraction of the tokens
that ranges over all values from 0 to 1~\cite{llada,illada}.
Left-to-right unmasking and right-to-left unmasking start from the token
sequence in which all tokens are masked.
The fraction of masked tokens then falls step by step from 1 to 0.
Every step of left-to-right unmasking and right-to-left unmasking therefore
lies within the range of the training of LLaDA-8B-Base and iLLaDA-8B-Base.
LLaDA-8B-Base and iLLaDA-8B-Base compute log-probabilities in bfloat16.

\paragraph{bert-base-uncased}
The training of bert-base-uncased masks 15\% of the tokens~\cite{bert}.
Left-to-right unmasking and right-to-left unmasking start far outside the
training mask ratio of bert-base-uncased, because all tokens are masked at the
first step.
Section~\ref{sec:mlmdelta} compares $\Delta_{\mathrm{block}}$ of
bert-base-uncased with $\Delta_{\mathrm{block}}$ of LLaDA-8B-Base and
iLLaDA-8B-Base in the light of the training mask ratio.
The tokenizer of bert-base-uncased splits a character string at spaces and
removes the spaces before tokenization, so that no token contains a space.
A word at the start of a character string and the same word after a space are
therefore the same token.
The tokenizer of bert-base-uncased therefore gives no mismatch with the
tokenizer output, with or without a leading space (Table~\ref{tab:mismatch}).
bert-base-uncased computes log-probabilities in 32-bit floating point.

\begin{table}[t]
\centering
\caption{Models of this paper.
The mask token is the token that marks a masked position.}
\label{tab:models}
\small
\begin{tabular}{llrl}
\hline
Model & Kind & Parameters & Quantities \\
\hline
gpt2 & autoregressive & 124 million & $\Delta_{\mathrm{block}}$ \\
gpt2-medium & autoregressive & 355 million & $\Delta_{\mathrm{block}}$ \\
gpt2-large & autoregressive & 774 million & $\Delta_{\mathrm{block}}$ \\
gpt2-xl & autoregressive & 1.5 billion & $\Delta_{\mathrm{block}}$ \\
Qwen3-4B-Base & autoregressive & 4.0 billion & $\Delta_{\mathrm{block}}$ \\
LLaDA-8B-Base & MLM (mask token 126336) & 8 billion & $\sigma$, $\Delta_{\mathrm{block}}$ \\
iLLaDA-8B-Base & MLM (mask token 5) & 8 billion & $\sigma$, $\Delta_{\mathrm{block}}$ \\
bert-base-uncased & MLM (mask token 103) & 110 million & $\sigma$, $\Delta_{\mathrm{block}}$ \\
\hline
\end{tabular}
\end{table}

\subsection{Reproduction of the values of Ref.~\cite{sagawa2026}}
\label{sec:reproduction}


Ref.~\cite{sagawa2026} publishes the block-level entropy production of every
written text of the five corpora of Ref.~\cite{sagawa2026}, together with the
number of tokens of every written text~\cite{sagawa2026}.
The published values exist for gpt2 and for Qwen3-4B-Base.
For each written text, this paper compares $\Delta_{\mathrm{block}}$ of the
written arrangement with the published block-level entropy production divided
by the published number of tokens.
Table~\ref{tab:reproduction} gives the comparison for the three settings of
Section~\ref{sec:tokenization}.

With setting 1, the largest absolute difference between
$\Delta_{\mathrm{block}}$ and the published value is $2.4\times10^{-5}$.
The number of tokens $L$ also equals the published number of tokens for all
1000 written texts.
Setting 1 therefore reproduces the values of Ref.~\cite{sagawa2026}.
With setting 2A, the median absolute difference between
$\Delta_{\mathrm{block}}$ and the published value is 0.25 to 0.37.
With setting 2B, the median absolute difference is 0.013 to 0.016.
With setting 2A and setting 2B, $L$ differs from the published number of tokens
for 71 of the 1000 written texts.
The difference in the number of tokens arises because the first word of the
written text can become a different number of tokens without a leading space.

For Qwen3-4B-Base, the comparison covers setting 1.
The largest absolute difference between $\Delta_{\mathrm{block}}$ and the
published value is $7.7\times10^{-3}$, and the median absolute difference is
$0.9\times10^{-3}$ to $1.2\times10^{-3}$ in each corpus.
The number of tokens $L$ also equals the published number of tokens for all
1000 written texts.
The size of the differences is judged against the values of
$\Delta_{\mathrm{block}}$ themselves.
For half of the 1000 written texts, the absolute value of
$\Delta_{\mathrm{block}}$ of the written arrangement exceeds 0.145.
The median absolute difference, about $1\times10^{-3}$, is therefore less than
1\% of a typical value of $\Delta_{\mathrm{block}}$.
The code of Ref.~\cite{sagawa2026} computes Qwen3-4B-Base in 16-bit floating
point on a GPU by default, and Ref.~\cite{sagawa2026} computed the values of
Qwen3-4B-Base on V100 GPUs~\cite[App.~C]{sagawa2026}.
This paper computed the values of Qwen3-4B-Base in 32-bit floating point.
The differences for Qwen3-4B-Base are consistent with the difference in
floating-point precision.
For gpt2, both Ref.~\cite{sagawa2026} and this paper computed in 32-bit
floating point, and the differences are below $3\times10^{-5}$.

\begin{table}[t]
\centering
\caption{Comparison of $\Delta_{\mathrm{block}}$ of the written arrangement,
computed with gpt2, with the block-level entropy production of
Ref.~\cite{sagawa2026} divided by the number of tokens of
Ref.~\cite{sagawa2026}, for the 200 written texts of each corpus of
Ref.~\cite{sagawa2026}.
For each setting, the table gives the largest and the median absolute
difference, and the number of written texts for which $L$ equals the number of
tokens of Ref.~\cite{sagawa2026}.
Each corpus carries the name of the large language model that generated the
corpus.
Fable5, GPT54, GPT56, Gemini31 and Opus46 stand for Claude Fable 5, GPT-5.4 Pro,
GPT-5.6 Pro, Gemini 3.1 Pro and Claude Opus 4.6.}
\label{tab:reproduction}
\small
\setlength{\tabcolsep}{4pt}
\begin{tabular}{lrrrrrr}
\hline
 & \multicolumn{2}{c}{Setting 1} & \multicolumn{2}{c}{Setting 2A} & \multicolumn{2}{c}{Setting 2B} \\
Corpus & max/median & $L$ & max/median & $L$ & max/median & $L$ \\
\hline
Fable5 & $1.9{\times}10^{-5}$/$4.9{\times}10^{-6}$ & 200 & 0.43/0.29 & 189 & 0.12/0.014 & 189 \\
GPT54 & $2.2{\times}10^{-5}$/$5.1{\times}10^{-6}$ & 200 & 0.50/0.37 & 200 & 0.071/0.016 & 200 \\
GPT56 & $2.4{\times}10^{-5}$/$4.7{\times}10^{-6}$ & 200 & 0.51/0.30 & 194 & 0.18/0.016 & 194 \\
Gemini31 & $2.4{\times}10^{-5}$/$4.5{\times}10^{-6}$ & 200 & 0.42/0.32 & 200 & 0.067/0.013 & 200 \\
Opus46 & $1.8{\times}10^{-5}$/$4.9{\times}10^{-6}$ & 200 & 0.43/0.25 & 146 & 0.15/0.016 & 146 \\
\hline
\end{tabular}
\end{table}

\subsection{Analyses of $\sigma$}
\label{sec:sigma-methods}


Section~\ref{sec:quantities} computes $\sigma$ for all 24 arrangements of each
written text, so that each written text has 24 values of $\sigma$.
Four analyses compare causal texts with non-causal texts.
The first analysis uses the midrank of the written arrangement among the 24
values of $\sigma$.
The second analysis uses the value of $\sigma$ for the written arrangement
only.
The third analysis and the fourth analysis use the 24 values of $\sigma$ to
locate the arrangement with the largest value of $\sigma$.

\paragraph{Midrank of the written arrangement}
For each written text, the midrank of the written arrangement among the 24
values of $\sigma$ follows Eq.~\eqref{eq:midrank}, with $\sigma(x_\pi)$ in
place of $\Delta_{\mathrm{block}}(\pi)$.
For each MLM and each corpus, the midrank is compared between causal texts and
non-causal texts.
The comparison reports the median midrank of each group, the rank-biserial
correlation $r$ and the $p$-value of the two-sided Mann--Whitney $U$ test.
For the corpora of Ref.~\cite{sagawa2026}, the comparison pools the five
corpora.
The authored component and the causal component of $\sigma$ both stay close to
zero, so that the causal share of $\sigma$ varies widely and carries no
information.
The causal share of $\sigma$ is therefore not reported.

\paragraph{Value of $\sigma$ for the written arrangement}
From the 24 values of each written text, the second analysis takes one value:
the value of $\sigma$ for the written arrangement.
The second analysis uses no arranged sentence character string and no other
arrangement.
The second analysis reports the mean value of $\sigma$ for causal texts minus
the mean value of $\sigma$ for non-causal texts.
The $p$-value comes from the Welch $t$ test.
For the corpora of Ref.~\cite{sagawa2026}, the comparison pools the five
corpora.

\paragraph{Cross table of the first sentence and the last sentence}
In the third analysis, the arrangement with the largest value of $\sigma$ is
located, for each written text, through the first sentence and the last
sentence of the arrangement.
A pair of a first sentence and a different last sentence is one of 12 pairs.
Each pair covers 2 of the 24 arrangements, because the two middle sentences
can appear in either order.
For each pair, the value of $\sigma$ is averaged over the 2 arrangements.
The written text is counted in the cell of the pair with the largest average.
The counts over the written texts of one group form a cross table.
The rows of the cross table are the 4 first sentences, and the columns are the
4 last sentences.
The 4 diagonal cells of the cross table are empty.
Without any effect of the order of the sentences on $\sigma$, each of the 12
cells has the probability $1/12$.
The test of the cross table compares the counts with the probability $1/12$
for each cell by the chi-square statistic.
The $p$-value of the test comes from 200\,000 random tables.
Each random table is drawn from the multinomial distribution with probability
$1/12$ for each cell and with the same number of written texts.
The $p$-value is the fraction of the random tables whose chi-square statistic
is at least the observed chi-square statistic.
Causal texts and non-causal texts are tested separately.

\paragraph{Test fixed before the computation for iLLaDA-8B-Base}
The fourth analysis is a test for iLLaDA-8B-Base.
The test was fixed on 26 November 2026, before the values of
$\sigma$ for iLLaDA-8B-Base with setting 1 were computed.
The direction of the hypotheses came from an earlier computation of $\sigma$
for iLLaDA-8B-Base without a leading space.
The test concerns the last sentence of the arrangement with the largest value
of $\sigma$ among the 6 arrangements that share one first sentence.
For a causal text and a first sentence among A, B and C, the last sentence is
one of three sentences, one of which is D.
Hypothesis 1 states that, for causal texts, the share of the cases in which
ending with D gives the largest value of $\sigma$ is below $1/3$.
Each case is a pair of a causal text and a first sentence among A, B and C.
For each of the three last sentences, the value of $\sigma$ is averaged over
the 2 arrangements with the first sentence and the last sentence.
Hypothesis 1 is tested by the one-sided exact binomial test, separately for the
Claude--Gemini corpus, the ChatGPT corpus and the corpora of
Ref.~\cite{sagawa2026}, and for all corpora pooled.
Hypothesis 2 states that, for non-causal texts, the share does not differ from
$1/3$, and that the share for causal texts is below the share for non-causal
texts.
The first part of hypothesis 2 is tested by the two-sided exact binomial test,
and the second part is tested by the one-sided Fisher exact test.
As a supplement, each causal text receives a count from 0 to 3.
The count is the number of first sentences among A, B and C for which ending
with D gives the largest value of $\sigma$.
The mean count over causal texts is compared with 1 by the one-sample $t$
test.

\section{Results}
\label{sec:results}

\subsection{$\Delta_{\mathrm{block}}$ with setting 1}
\label{sec:delta}


\paragraph{Corpora of Ref.~\cite{sagawa2026}}
Each of the five autoregressive models of Table~\ref{tab:models} computed
$\Delta_{\mathrm{block}}$ with setting 1 for the written texts of the corpora of
Ref.~\cite{sagawa2026}.
Table~\ref{tab:share-sagawa} gives the authored component, the causal component
and the causal share of $\Delta_{\mathrm{block}}$ for each of the five
autoregressive models.
The causal share is 71\% to 93\%, and the causal share in $z$ is 73\% to 92\%.
The authored component is $+0.021$ to $+0.104$, and the causal component is
$+0.187$ to $+0.292$.
Over the five models and the five corpora, the median midrank of the written
arrangement is 0.48 to 0.69 for non-causal texts and 0.65 to 0.90 for causal
texts.
The separation depends on the corpus.
For gpt2-medium, gpt2-xl and Qwen3-4B-Base, the rank-biserial correlation $r$
of the midrank between causal texts and non-causal texts exceeds 0.2 in all 5
corpora.
For gpt2 and gpt2-large, $r$ exceeds 0.2 in 3 of the 5 corpora.
In the two other corpora, which GPT-5.4 Pro and GPT-5.6 Pro generated, $r$ is
$+0.09$ to $+0.15$.
For gpt2, Ref.~\cite{sagawa2026} also finds a weaker separation in the same
two corpora and relates the weaker separation to properties of the large
language model that generated the corpus~\cite[Sec.~V~B]{sagawa2026}.
The present results do not determine whether the weaker separation comes from
a similarity between the causal texts and the non-causal texts of a corpus.

\begin{table}[t]
\centering
\caption{Authored component, causal component and causal share of
$\Delta_{\mathrm{block}}$ with setting 1 on the five corpora of
Ref.~\cite{sagawa2026}.
The last column counts the corpora of Ref.~\cite{sagawa2026} in which the
rank-biserial correlation $r$ of the midrank between causal texts and
non-causal texts exceeds 0.2.}
\label{tab:share-sagawa}
\small
\setlength{\tabcolsep}{4pt}
\begin{tabular}{lrrrrr}
\hline
 & Authored & Causal & Causal & Causal share & Corpora \\
Model & component & component & share & in $z$ & with $r > 0.2$ \\
\hline
gpt2 & $+0.062$ & $+0.187$ & 75\% & 82\% & 3 of 5 \\
gpt2-medium & $+0.062$ & $+0.208$ & 77\% & 83\% & 5 of 5 \\
gpt2-large & $+0.062$ & $+0.250$ & 80\% & 75\% & 3 of 5 \\
gpt2-xl & $+0.104$ & $+0.250$ & 71\% & 73\% & 5 of 5 \\
Qwen3-4B-Base & $+0.021$ & $+0.292$ & 93\% & 92\% & 5 of 5 \\
\hline
\end{tabular}
\end{table}

\paragraph{Claude--Gemini corpus and ChatGPT corpus}
Table~\ref{tab:share-graph} gives the same quantities for the Claude--Gemini
corpus and the ChatGPT corpus.
On the Claude--Gemini corpus, the causal share is 64\% to 100\%, and $r$ of the
midrank is $+0.14$ to $+0.27$ ($p \le 1.5\times10^{-3}$).
On the ChatGPT corpus, the causal share is 62\% to 127\%, and $r$ of the
midrank is $+0.33$ to $+0.46$ ($p \le 3.8\times10^{-4}$).
The causal share exceeds 100\% for gpt2 and Qwen3-4B-Base on the ChatGPT
corpus, because the authored component is negative.
In the causal texts of the Claude--Gemini corpus and the ChatGPT corpus, the
causal graph fixed before generation and checked by two judge models
determines the causal order.
$\Delta_{\mathrm{block}}$ therefore separates causal texts from non-causal texts
also when causal texts are defined by a causal graph and not by a generation
prompt alone.

\begin{table}[t]
\centering
\caption{$\Delta_{\mathrm{block}}$ with setting 1 on the Claude--Gemini corpus
(338 causal texts, 350 non-causal texts) and the ChatGPT corpus (79 causal
texts, 80 non-causal texts).
Midrank: median midrank of the written arrangement for causal texts and
non-causal texts, and $r$ between the two groups.
Qwen3-4B stands for Qwen3-4B-Base.
Written arrangement: median value of $\Delta_{\mathrm{block}}$ for the written
arrangement for causal texts and non-causal texts, and $r$ between the two
groups.}
\label{tab:share-graph}
\small
\setlength{\tabcolsep}{3pt}
\begin{tabular}{lrrrrrr}
\hline
 & \multicolumn{2}{c}{Midrank} & Causal & \multicolumn{3}{c}{Written arrangement} \\
Model & causal/non-causal & $r$ & share & causal & non-causal & $r$ \\
\hline
\multicolumn{7}{l}{Claude--Gemini corpus} \\
\quad gpt2 & 0.604/0.500 & $+0.14$ & 100\% & $+0.029$ & $-0.000$ & $+0.13$ \\
\quad gpt2-medium & 0.688/0.563 & $+0.19$ & 67\% & $+0.050$ & $+0.015$ & $+0.21$ \\
\quad gpt2-large & 0.729/0.563 & $+0.27$ & 73\% & $+0.062$ & $+0.009$ & $+0.28$ \\
\quad gpt2-xl & 0.729/0.583 & $+0.23$ & 64\% & $+0.075$ & $+0.018$ & $+0.26$ \\
\quad Qwen3-4B & 0.729/0.521 & $+0.27$ & 91\% & $+0.075$ & $+0.002$ & $+0.24$ \\
\hline
\multicolumn{7}{l}{ChatGPT corpus} \\
\quad gpt2 & 0.729/0.438 & $+0.33$ & 127\% & $+0.059$ & $-0.014$ & $+0.31$ \\
\quad gpt2-medium & 0.813/0.521 & $+0.46$ & 93\% & $+0.123$ & $+0.006$ & $+0.52$ \\
\quad gpt2-large & 0.813/0.521 & $+0.41$ & 93\% & $+0.097$ & $+0.007$ & $+0.43$ \\
\quad gpt2-xl & 0.771/0.604 & $+0.33$ & 62\% & $+0.099$ & $+0.028$ & $+0.38$ \\
\quad Qwen3-4B & 0.771/0.479 & $+0.45$ & 108\% & $+0.112$ & $-0.012$ & $+0.38$ \\
\hline
\end{tabular}
\end{table}

\paragraph{Value of $\Delta_{\mathrm{block}}$ for the written arrangement}
Ref.~\cite{sagawa2026} compares causal texts with non-causal texts through the
block-level entropy production of the written arrangement, without the other
arrangements~\cite[Sec.~V~B]{sagawa2026}.
This paper makes the same comparison on the Claude--Gemini corpus and the
ChatGPT corpus, with the value of $\Delta_{\mathrm{block}}$ for the written
arrangement.
The last three columns of Table~\ref{tab:share-graph} give the comparison.
For non-causal texts, the median value of $\Delta_{\mathrm{block}}$ for the
written arrangement lies between $-0.014$ and $+0.028$, close to zero.
For causal texts, the median value lies between $+0.029$ and $+0.123$.
The comparison of the values for the written arrangement thus agrees with the
comparison of the midrank.

\subsection{$\Delta_{\mathrm{block}}$ with setting 2A and setting 2B}
\label{sec:settings}


\paragraph{Setting 2A}
Table~\ref{tab:settings} gives the causal share of $\Delta_{\mathrm{block}}$
with the three settings.
With setting 2A, the causal share drops to 10\% to 19\% on the corpora of
Ref.~\cite{sagawa2026}, to 0\% to 19\% on the Claude--Gemini corpus and to 10\%
to 21\% on the ChatGPT corpus.
With setting 2A, the authored component rises to $+0.31$ to $+0.40$, whereas
the authored component with setting 1 is $-0.06$ to $+0.10$.
With setting 2A, every arrangement that does not start with sentence A has two
mismatches with the tokenizer output (Table~\ref{tab:mismatch}).
The written arrangement starts with sentence A and has no mismatch.
The rise of the authored component is consistent with a higher rank of the
arrangements without a mismatch, in causal texts and in non-causal texts alike.
The first version of this paper~\cite{taguchi2026v1} computed
$\Delta_{\mathrm{block}}$ with setting 2A, and the values of the first version
for the corpora of Ref.~\cite{sagawa2026} are the values of setting 2A in
Table~\ref{tab:settings}.

\paragraph{Setting 2B}
With setting 2B, the causal share is 62\% to 93\% on the corpora of
Ref.~\cite{sagawa2026}, 67\% to 108\% on the Claude--Gemini corpus and 77\% to
111\% on the ChatGPT corpus.
The causal share with setting 2B stays close to the causal share with
setting 1.
Setting 2B has no mismatch with the tokenizer output, but setting 2B does not
satisfy the requirement of the definition of Ref.~\cite{sagawa2026}
(Section~\ref{sec:tokenization}).

\begin{table}[t]
\centering
\caption{Causal share of $\Delta_{\mathrm{block}}$ with the three settings.
Setting 1: procedure 2-T with a leading space.
Setting 2A: procedure 2-T without a leading space.
Setting 2B: procedure 2-S without a leading space.}
\label{tab:settings}
\small
\setlength{\tabcolsep}{4pt}
\begin{tabular}{lrrr}
\hline
Model & Setting 1 & Setting 2A & Setting 2B \\
\hline
\multicolumn{4}{l}{Corpora of Ref.~\cite{sagawa2026}} \\
\quad gpt2 & 75\% & 11\% & 62\% \\
\quad gpt2-medium & 77\% & 19\% & 77\% \\
\quad gpt2-large & 80\% & 11\% & 80\% \\
\quad gpt2-xl & 71\% & 10\% & 80\% \\
\quad Qwen3-4B-Base & 93\% & 10\% & 93\% \\
\hline
\multicolumn{4}{l}{Claude--Gemini corpus} \\
\quad gpt2 & 100\% & 11\% & 86\% \\
\quad gpt2-medium & 67\% & 0\% & 67\% \\
\quad gpt2-large & 73\% & 11\% & 91\% \\
\quad gpt2-xl & 64\% & 10\% & 73\% \\
\quad Qwen3-4B-Base & 91\% & 19\% & 108\% \\
\hline
\multicolumn{4}{l}{ChatGPT corpus} \\
\quad gpt2 & 127\% & 21\% & 111\% \\
\quad gpt2-medium & 93\% & 19\% & 87\% \\
\quad gpt2-large & 93\% & 19\% & 92\% \\
\quad gpt2-xl & 62\% & 10\% & 77\% \\
\quad Qwen3-4B-Base & 108\% & 19\% & 107\% \\
\hline
\end{tabular}
\end{table}

\subsection{$\sigma$ of three MLMs}
\label{sec:sigma}


\paragraph{Midrank of the written arrangement}
Table~\ref{tab:sigma} gives the results of the first three analyses of
Section~\ref{sec:sigma-methods} for LLaDA-8B-Base, iLLaDA-8B-Base and
bert-base-uncased.
The median midrank of the written arrangement among the 24 values of $\sigma$
lies between 0.39 and 0.61 for causal texts and for non-causal texts.
The rank-biserial correlation $r$ of the midrank lies between $-0.15$ and
$+0.12$.
The sign of $r$ changes with the MLM and with the corpus.
In contrast, $r$ of the midrank of $\Delta_{\mathrm{block}}$ with setting 1 lies
between $+0.14$ and $+0.46$ (Table~\ref{tab:share-graph}).
The range covers every autoregressive model on the Claude--Gemini corpus and
the ChatGPT corpus.

\paragraph{Value of $\sigma$ for the written arrangement}
The mean value of $\sigma$ for the written arrangement in causal texts minus the
mean value in non-causal texts lies between $-0.045$ and $+0.034$ for
LLaDA-8B-Base and iLLaDA-8B-Base.
For bert-base-uncased, the difference is $-0.015$ on the Claude--Gemini corpus,
$+0.140$ on the ChatGPT corpus and $-0.322$ on the corpora of
Ref.~\cite{sagawa2026}.
The sign of the difference changes with the corpus for all three MLMs.

\paragraph{Cross table of the first sentence and the last sentence}
The midrank and the value of $\sigma$ for the written arrangement show no
consistent separation of causal texts from non-causal texts.
The cross table checks whether $\sigma$ responds to the order of the sentences
at all.
The result shows that $\sigma$ responds to the order of the sentences in some
groups of causal texts, but not strongly enough to separate causal texts from
non-causal texts.
For causal texts of the Claude--Gemini corpus, the test of the cross table gives
$p = 0.0019$ for LLaDA-8B-Base, $p = 0.017$ for iLLaDA-8B-Base and
$p = 0.0018$ for bert-base-uncased.
For non-causal texts of the Claude--Gemini corpus, the test gives $p \ge 0.14$
for the three MLMs.
For causal texts of the corpora of Ref.~\cite{sagawa2026}, the test gives
$p = 0.069$ for LLaDA-8B-Base, $p = 1.5\times10^{-4}$ for iLLaDA-8B-Base and
$p = 0.038$ for bert-base-uncased.
For non-causal texts of the corpora of Ref.~\cite{sagawa2026}, the test gives
$p = 0.33$ for LLaDA-8B-Base, $p = 0.014$ for iLLaDA-8B-Base and $p = 0.19$ for
bert-base-uncased.
The three MLMs and the three groups of corpora give 9 groups of causal texts
and 9 groups of non-causal texts.
The test gives $p < 0.05$ in 5 of the 9 groups of causal texts and in 1 of the
9 groups of non-causal texts.
The order of the sentences therefore affects $\sigma$ mainly in causal texts,
and rarely in non-causal texts.
The effect in causal texts is not consistent across the corpora: the test gives
$p \ge 0.13$ for all three MLMs on the ChatGPT corpus.

\begin{table}[t]
\centering
\caption{$\sigma$ of three MLMs.
Midrank: median midrank of the written arrangement among the 24 values of
$\sigma$ for causal texts (caus.) and non-causal texts (non-caus.), and $r$
between the two groups.
Written arrangement: mean value of $\sigma$ for the written arrangement in
causal texts minus the mean value in non-causal texts, and the $p$-value of
the Welch $t$ test.
Cross table: $p$-value of the test of the cross table for causal texts and for
non-causal texts.
LLaDA, iLLaDA and bert stand for LLaDA-8B-Base, iLLaDA-8B-Base and
bert-base-uncased.}
\label{tab:sigma}
\small
\setlength{\tabcolsep}{2.5pt}
\begin{tabular}{lrrrrrr}
\hline
 & \multicolumn{2}{c}{Midrank} & \multicolumn{2}{c}{Written arrangement} & \multicolumn{2}{c}{Cross table} \\
MLM & caus./non-caus. & $r$ & difference & $p$ & causal & non-causal \\
\hline
\multicolumn{7}{l}{Claude--Gemini corpus} \\
\quad LLaDA & 0.479/0.438 & $+0.05$ & $+0.019$ & $1.5\times10^{-1}$ & $1.9\times10^{-3}$ & 0.63 \\
\quad iLLaDA & 0.479/0.479 & $-0.05$ & $+0.021$ & $5.0\times10^{-2}$ & $1.7\times10^{-2}$ & 0.14 \\
\quad bert & 0.583/0.479 & $+0.12$ & $-0.015$ & $5.1\times10^{-1}$ & $1.8\times10^{-3}$ & 0.39 \\
\hline
\multicolumn{7}{l}{ChatGPT corpus} \\
\quad LLaDA & 0.438/0.521 & $-0.07$ & $-0.017$ & $5.0\times10^{-1}$ & $8.1\times10^{-1}$ & 0.34 \\
\quad iLLaDA & 0.396/0.542 & $-0.15$ & $+0.034$ & $1.4\times10^{-1}$ & $3.3\times10^{-1}$ & 0.49 \\
\quad bert & 0.604/0.521 & $+0.12$ & $+0.140$ & $1.3\times10^{-3}$ & $1.3\times10^{-1}$ & 0.16 \\
\hline
\multicolumn{7}{l}{Corpora of Ref.~\cite{sagawa2026}} \\
\quad LLaDA & 0.438/0.521 & $-0.10$ & $-0.045$ & $1.4\times10^{-4}$ & $6.9\times10^{-2}$ & 0.33 \\
\quad iLLaDA & 0.438/0.521 & $-0.08$ & $-0.013$ & $2.4\times10^{-1}$ & $1.5\times10^{-4}$ & 0.014 \\
\quad bert & 0.438/0.521 & $-0.13$ & $-0.322$ & $1.6\times10^{-44}$ & $3.8\times10^{-2}$ & 0.19 \\
\hline
\end{tabular}
\end{table}

\paragraph{Test fixed before the computation for iLLaDA-8B-Base}
Table~\ref{tab:prereg} gives the result of the fourth analysis of
Section~\ref{sec:sigma-methods}.
For causal texts, the share of the cases in which ending with D gives the
largest value of $\sigma$ is 0.295 to 0.312 in the three groups of corpora.
The share is smaller than $1/3$ in all three groups.
For all corpora pooled, the share is 846 of 2751 cases, 0.308, and the
one-sided exact binomial test gives $p = 0.0021$.
Hypothesis 1 is therefore supported for the pooled corpora, with a small but
statistically significant effect: 0.308 against $1/3$.
The supplementary test also supports hypothesis 1: the mean count is 0.923,
and the one-sided $t$ test against 1 gives $p = 0.012$.
For non-causal texts, the share is 0.341 and 0.336 on the Claude--Gemini corpus
and the corpora of Ref.~\cite{sagawa2026}, close to $1/3$.
On the ChatGPT corpus, the share for non-causal texts is 0.438, larger than
$1/3$ ($p = 7.7\times10^{-4}$), against the first part of hypothesis 2.
The share for causal texts is below the share for non-causal texts for all
corpora pooled ($p = 0.0011$, Fisher exact test).

\begin{table}[t]
\centering
\caption{Test fixed before the computation for iLLaDA-8B-Base.
The table gives the share of the cases in which ending with D gives the largest
value of $\sigma$.
A case is a pair of a written text and a first sentence among A, B and C.
Chance: $1/3$.
Causal texts: one-sided exact binomial test (share below $1/3$).
Non-causal texts: two-sided exact binomial test.
Fisher: one-sided Fisher exact test (share for causal texts below the share for
non-causal texts).}
\label{tab:prereg}
\small
\begin{tabular}{lrrrrr}
\hline
 & \multicolumn{2}{c}{Causal texts} & \multicolumn{2}{c}{Non-causal texts} & \\
Corpus & share & $p$ & share & $p$ & Fisher $p$ \\
\hline
Claude--Gemini & 316/1014 = 0.312 & 0.075 & 0.341 & $6.0\times10^{-1}$ & $8.5\times10^{-2}$ \\
ChatGPT & 70/237 = 0.295 & 0.12 & 0.438 & $7.7\times10^{-4}$ & $8.7\times10^{-4}$ \\
Ref.~\cite{sagawa2026} & 460/1500 = 0.307 & 0.015 & 0.336 & $8.3\times10^{-1}$ & $4.6\times10^{-2}$ \\
Pooled & 846/2751 = 0.308 & 0.0021 & 0.347 & $1.4\times10^{-1}$ & $1.1\times10^{-3}$ \\
\hline
\end{tabular}
\end{table}

\paragraph{Summary for $\sigma$}
The three analyses that do not depend on a hypothesis fixed in advance show no
consistent separation of causal texts from non-causal texts by $\sigma$.
The test fixed before the computation finds a small but statistically
significant effect in the direction of hypothesis 1 for the pooled corpora.
The value of $\sigma$ thus carries weak information about the order of the
sentences, and the information is much weaker than the information in
$\Delta_{\mathrm{block}}$.

\subsection{$\Delta_{\mathrm{block}}$ of three MLMs}
\label{sec:mlmdelta}


\paragraph{LLaDA-8B-Base and iLLaDA-8B-Base}
The three MLMs of Section~\ref{sec:sigma} also computed
$\Delta_{\mathrm{block}}$ with left-to-right unmasking
(Eq.~\eqref{eq:dblock-l2r}) and setting 1.
The computation reuses the log-probabilities $\log P_{\mathrm{l2r}}(x_\pi)$ of
the 24 arrangements that the computation of $\sigma$ recorded.
Table~\ref{tab:mlmdelta} gives the results.
For LLaDA-8B-Base and iLLaDA-8B-Base, the rank-biserial correlation $r$ of the
midrank between causal texts and non-causal texts is $+0.35$ to $+0.57$
($p \le 2.5\times10^{-8}$).
The causal component is $+0.21$ to $+0.42$, and the causal share is 77\% to
133\%.
For non-causal texts, the median value of $\Delta_{\mathrm{block}}$ for the
written arrangement lies between $-0.033$ and $+0.019$, close to zero.
For causal texts, the median value lies between $+0.060$ and $+0.136$.
$\Delta_{\mathrm{block}}$ of LLaDA-8B-Base and iLLaDA-8B-Base thus separates
causal texts from non-causal texts as clearly as $\Delta_{\mathrm{block}}$ of
the autoregressive models (Tables~\ref{tab:share-sagawa}
and~\ref{tab:share-graph}).
The values of $\sigma$ of the same two MLMs separate causal texts from
non-causal texts only weakly (Section~\ref{sec:sigma}).

\paragraph{bert-base-uncased}
For bert-base-uncased, $r$ of the midrank is $+0.14$ ($p = 1.1\times10^{-3}$)
on the Claude--Gemini corpus and $+0.10$ ($p = 0.29$) on the ChatGPT corpus.
On the corpora of Ref.~\cite{sagawa2026}, $r$ of the midrank is $-0.05$
($p = 0.19$).
The causal component of bert-base-uncased is $-0.04$ to $+0.08$, close to
zero, so that the causal share of bert-base-uncased varies widely and carries
no information.
$\Delta_{\mathrm{block}}$ of bert-base-uncased therefore separates causal texts
from non-causal texts weakly or not at all.

\paragraph{Training mask ratio}
The difference between bert-base-uncased and the two other MLMs is consistent
with the difference in the training mask ratio (Section~\ref{sec:models}).
Left-to-right unmasking starts from the token sequence in which all tokens are
masked.
The training of LLaDA-8B-Base and iLLaDA-8B-Base covers every fraction of masked
tokens, so that every step of left-to-right unmasking lies within the range of
the training.
The training of bert-base-uncased masks 15\% of the tokens.
The log-probability $\log P_{\mathrm{l2r}}$ of bert-base-uncased therefore
rests on predictions far outside the training of bert-base-uncased.

\begin{table}[t]
\centering
\caption{$\Delta_{\mathrm{block}}$ of three MLMs with left-to-right unmasking
and setting 1.
Midrank: median midrank of the written arrangement for causal texts (caus.) and
non-causal texts (non-caus.), and $r$ between the two groups.
Written arrangement: median value of $\Delta_{\mathrm{block}}$ for the written
arrangement for causal texts and non-causal texts.
LLaDA, iLLaDA and bert stand for LLaDA-8B-Base, iLLaDA-8B-Base and
bert-base-uncased.
The causal share of bert-base-uncased is omitted, because the causal component
of bert-base-uncased is close to zero.}
\label{tab:mlmdelta}
\small
\setlength{\tabcolsep}{3pt}
\begin{tabular}{lrrrrrr}
\hline
 & \multicolumn{2}{c}{Midrank} & Causal & Causal & \multicolumn{2}{c}{Written arrangement} \\
MLM & caus./non-caus. & $r$ & component & share & causal & non-causal \\
\hline
\multicolumn{7}{l}{Claude--Gemini corpus} \\
\quad LLaDA & 0.729/0.479 & $+0.37$ & $+0.250$ & 109\% & $+0.072$ & $-0.008$ \\
\quad iLLaDA & 0.729/0.521 & $+0.35$ & $+0.208$ & 91\% & $+0.060$ & $+0.001$ \\
\quad bert & 0.562/0.479 & $+0.14$ & $+0.083$ & -- & $+0.042$ & $-0.011$ \\
\hline
\multicolumn{7}{l}{ChatGPT corpus} \\
\quad LLaDA & 0.854/0.562 & $+0.57$ & $+0.292$ & 82\% & $+0.125$ & $+0.010$ \\
\quad iLLaDA & 0.812/0.396 & $+0.51$ & $+0.417$ & 133\% & $+0.094$ & $-0.033$ \\
\quad bert & 0.562/0.562 & $+0.10$ & $+0.000$ & -- & $+0.026$ & $+0.020$ \\
\hline
\multicolumn{7}{l}{Corpora of Ref.~\cite{sagawa2026}} \\
\quad LLaDA & 0.854/0.562 & $+0.45$ & $+0.250$ & 80\% & $+0.136$ & $+0.019$ \\
\quad iLLaDA & 0.812/0.521 & $+0.41$ & $+0.208$ & 77\% & $+0.119$ & $+0.015$ \\
\quad bert & 0.479/0.521 & $-0.05$ & $-0.042$ & -- & $-0.018$ & $+0.008$ \\
\hline
\end{tabular}
\end{table}

\section{Discussion}
\label{sec:discussion}


\subsection{Dependence of $\Delta_{\mathrm{block}}$ on the leading space}
\label{sec:fragility}

The definition of the block-level entropy production in Ref.~\cite{sagawa2026}
requires procedure 2-T (Section~\ref{sec:tokenization}).
Within procedure 2-T, setting 1 has no mismatch with the tokenizer output, and
setting 2A has two mismatches in every arrangement that does not start with
sentence A.
The definition of $\Delta_{\mathrm{block}}$ does not state whether a leading
space is added.
The choice of the leading space nevertheless reverses the conclusion.
On the corpora of Ref.~\cite{sagawa2026}, the causal share is 71\% to 93\% with
setting 1 and 10\% to 19\% with setting 2A (Table~\ref{tab:settings}).
Setting 1 and setting 2A differ at two positions of each arrangement that does
not start with sentence A.
The token sequence of a written text has 26 to 82 tokens, with a median of 59
tokens for gpt2 and 45 tokens for Qwen3-4B-Base.
A conclusion drawn from $\Delta_{\mathrm{block}}$ therefore depends on a choice
that the definition of $\Delta_{\mathrm{block}}$ does not fix.

Ref.~\cite{sagawa2026} adds the leading space and states the purpose of the
leading space: the leading space preserves the space between sentences when the
sentence token sequences are rearranged~\cite[App.~B~4 and
App.~C]{sagawa2026}.
Ref.~\cite{sagawa2026} does not report how much the leading space changes the
comparison between causal texts and non-causal texts.
Table~\ref{tab:settings} shows that the leading space decides the sign of the
conclusion.

Setting 2B gives a causal share close to the causal share of setting 1
(Table~\ref{tab:settings}).
Setting 2B does not satisfy the requirement of the definition of
Ref.~\cite{sagawa2026}, because the first token of sentence A depends on the
position of sentence A.
Setting 2B therefore cannot replace setting 1 in a computation of the
block-level entropy production.
A model that receives the token sequences of setting 2A receives token
sequences that the tokenizer never produces.
Within the definition of Ref.~\cite{sagawa2026}, however, the only way to avoid
the token sequences of setting 2A is the leading space, which lies outside the
definition.

\subsection{The first version of this paper}
\label{sec:firstversion}

The first version of this paper~\cite{taguchi2026v1} computed
$\Delta_{\mathrm{block}}$ with setting 2A.
Setting 2A satisfies the requirement of the definition of
Ref.~\cite{sagawa2026}, and the values of the first version are values of
$\Delta_{\mathrm{block}}$.
The first version did not recognize that the choice of the leading space can
reverse the conclusion.
The first version reported a causal share of 8\% to 19\%.
The first version concluded that the written arrangement, and not the causal
order, explains most of the high midrank of the written arrangement.
With setting 1, the causal share is 71\% to 93\%, and the conclusion of the
first version is reversed.
The first version of this paper has been withdrawn.

\subsection{Separation of causal texts and entropy production}
\label{sec:separation}

Ref.~\cite{sagawa2026} finds larger values of the block-level entropy
production for causal texts.
Ref.~\cite{sagawa2026} reads the finding as a first step toward probing the
temporal order or the causal order of the events that a text describes.
The results of Section~\ref{sec:results} allow a comparison of two quantities
computed by the same MLMs.

The quantity $\sigma$ compares left-to-right unmasking and right-to-left
unmasking of the same token sequence.
For an MLM, left-to-right unmasking and right-to-left unmasking are two
processes that the MLM actually runs on the same token sequence, in opposite
unmasking orders.
For an MLM whose conditional probabilities come from one joint distribution,
$\sigma$ is zero for every token sequence.
The quantity $\sigma$ therefore has a zero on each token sequence, as an
entropy production has a zero for a reversible process.
The values of $\sigma$ of LLaDA-8B-Base and iLLaDA-8B-Base separate causal texts
from non-causal texts only weakly (Section~\ref{sec:sigma}).

The quantity $\Delta_{\mathrm{block}}$ compares the token sequences of two
arrangements.
The relation $\Delta_{\mathrm{block}}(\bar{\pi}) = -\Delta_{\mathrm{block}}(\pi)$
makes the 24 values of one written text symmetric about zero.
However, no theorem fixes the value of $\Delta_{\mathrm{block}}$ that
corresponds to no irreversibility for one written text~\cite{taguchi2026v1}.
The quantity $\Delta_{\mathrm{block}}$ therefore has no zero that holds on each
written text.
$\Delta_{\mathrm{block}}$ of the same two MLMs separates causal texts from
non-causal texts clearly (Section~\ref{sec:mlmdelta}).

With respect to the zero, $\sigma$ is closer to an entropy production than
$\Delta_{\mathrm{block}}$.
The quantity closer to an entropy production loses the separation, and the
quantity that keeps the separation is a comparison between two token
sequences.
The ability of a quantity to separate causal texts from non-causal texts is
therefore not evidence that the quantity is an entropy production.

\section{Limitations}
\label{sec:limitations}


\paragraph{Causal order and temporal order}
The comparison of this paper sets causal texts against non-causal texts.
In causal texts, the content fixes the order of the sentences through a chain
of causes.
In non-causal texts, nothing fixes the order of the sentences.
This paper uses no texts in which the events follow a temporal order without a
causal relation.
The results therefore do not determine whether $\Delta_{\mathrm{block}}$ responds
to causal order itself or to any order that the content fixes.
The question corresponds to the second open problem of
Ref.~\cite[Sec.~V~B and Sec.~VIII]{sagawa2026}.

\paragraph{Generation and judging by large language models}
Large language models generated the written texts of all corpora.
Two judge models, not human readers, checked the causal graph of each written
text of the Claude--Gemini corpus and the ChatGPT corpus.
The judge models came from companies whose large language models did not
generate the written text, but the judge models and the generating large
language models may still share biases.
Each batch of 40 written texts came from one response of one large language
model.
The written texts within one batch are therefore not guaranteed to be
statistically independent.
Ref.~\cite[App.~B~4]{sagawa2026} notes the same point for the corpora of
Ref.~\cite{sagawa2026}.

\paragraph{Form of the written texts}
Every written text consists of four short English sentences.
The results may not carry over to longer texts, to texts with a different
number of sentences, or to other languages.

\paragraph{Weaker separation in some corpora}
For gpt2 and gpt2-large, the rank-biserial correlation $r$ of the midrank of
$\Delta_{\mathrm{block}}$ with setting 1 is at most 0.2 in the two corpora of
Ref.~\cite{sagawa2026} that GPT-5.4 Pro and GPT-5.6 Pro generated
(Section~\ref{sec:delta}).
The weaker separation may come from a similarity between the causal texts and
the non-causal texts of a corpus.
The weaker separation may also come from properties of the large language model
that generated the corpus.
The results do not distinguish between the two possibilities.

\paragraph{Origin of the hypotheses for iLLaDA-8B-Base}
The test for iLLaDA-8B-Base was fixed before the computation of $\sigma$ with
setting 1 (Section~\ref{sec:sigma-methods}).
The direction of the hypotheses, however, came from an earlier computation of
$\sigma$ for iLLaDA-8B-Base without a leading space.
The earlier computation had mismatches with the tokenizer output, and the
earlier computation is not part of the results of this paper.

\section{Conclusions}
\label{sec:conclusions}


This paper evaluated the quantity $\Delta_{\mathrm{block}}$ of
Ref.~\cite{sagawa2026}, whose value for the written arrangement times $L$ is
the block-level entropy production, over the 24 arrangements of the four
sentences of each written text.
With setting 1, which reproduces the values of Ref.~\cite{sagawa2026}, the
causal share is 71\% to 93\% on the corpora of Ref.~\cite{sagawa2026}.
$\Delta_{\mathrm{block}}$ also separates causal texts from non-causal texts on
two corpora whose causal texts are defined by causal graphs fixed before
generation and checked by judge models.
The separation, however, depends on a choice outside the definition of
$\Delta_{\mathrm{block}}$: without a leading space, the causal share drops to
0\% to 21\%.
The masked language models LLaDA-8B-Base and iLLaDA-8B-Base give two
quantities for the same written texts.
The quantity $\sigma$ has a zero on each token sequence and separates causal
texts from non-causal texts only weakly.
The quantity $\Delta_{\mathrm{block}}$ has no zero that holds on each written
text and separates causal texts from non-causal texts clearly.
The ability of a quantity to separate causal texts from non-causal texts is
therefore not evidence that the quantity is an entropy production.

\section*{Acknowledgements}
Computation time was provided by the SuperComputer System, Institute for
Chemical Research, Kyoto University.

\section*{Declaration of competing interest}
The author declares no competing interests.

\section*{Data availability}
The corpora, the computed values and the code that produces every number of
this paper are available at \url{https://github.com/tagtag/sigma-order}.
The repository becomes public when the preprint of this paper is announced on
arXiv, and the version used for this paper will be archived on Zenodo with a
DOI.

\section*{Declaration of generative AI and AI-assisted technologies in the writing process}
During the preparation of this work the author used Claude Opus 5.5
(Anthropic) in order to draft the English text from claims, terms and
results that the author fixed in Japanese, to revise the text, and to write
and run the analysis and verification scripts.
The author decided every claim and every term of the paper, and checked every
sentence, every number and every script.
After using this tool, the author reviewed and edited the content as needed
and takes full responsibility for the content of the publication.
The use of large language models in the research itself, for the generation
and the judging of the corpora, is described in Section~\ref{sec:corpus}.


\bibliographystyle{elsarticle-num}
\bibliography{refs}


\end{document}